\documentclass[conference]{IEEEtran}
\IEEEoverridecommandlockouts
\usepackage{amsmath,amsfonts}
\usepackage{algpseudocode}
\usepackage{array}
\usepackage{booktabs} 
\usepackage[caption=false,font=normalsize,labelfont=sf,textfont=sf]{subfig}
\usepackage{textcomp}
\usepackage{stfloats}
\usepackage{url}
\usepackage{verbatim}
\usepackage{graphicx}
\usepackage{cite}
\usepackage{acronym}
\usepackage{float}
\usepackage{courier}

\usepackage{ifpdf}
\usepackage{url}
\ifpdf
\else
\fi

\usepackage{cite}
\usepackage{balance}
\usepackage{bm,comment,color}

\usepackage{amsmath}

\usepackage{array}
\usepackage{amsfonts} 
\usepackage{amssymb}
\usepackage{esint} % various fancy integral symbols
\usepackage{units}
\usepackage{multirow}
\usepackage{comment}
\usepackage{url}
\usepackage{xcolor}

\usepackage{color}
\definecolor{lightbrown}{RGB}{210,180,140}

\usepackage{pgfplots}
\usepackage{tikz}
\usetikzlibrary{calc}
\makeatletter
\newcommand{\gettikzxy}[3]{%
  \tikz@scan@one@point\pgfutil@firstofone#1\relax
  \edef#2{\the\pgf@x}%
  \edef#3{\the\pgf@y}%
}
\usetikzlibrary{spy,backgrounds}

\pgfplotsset{compat=newest}
\usetikzlibrary{plotmarks}
\usetikzlibrary{arrows.meta}
\usepgfplotslibrary{patchplots}
\usepackage{grffile}
\newlength\fheight 
\newlength\fwidth 
\usepgfplotslibrary{fillbetween}

\allowdisplaybreaks

\acrodef{ad}[AD]{autonomous drive}
\acrodef{adas}[ADAS]{advanced driver assistance system}
\acrodef{aoa}[AOA]{angles-of-arrival}
\acrodef{aod}[AOD]{angles-of-departure}\acrodef{bs}[BS]{base station}
\acrodef{cdf}[CDF]{cumulative distribution function}
\acrodef{cir}[CIR]{channel impulse response}
\acrodef{cfr}[CFR]{channel frequency response}
\acrodef{dbscan}[DBSCAN]{density-based spatial clustering of applications with noise}
\acrodef{dt}[DT]{digital twin}
\acrodef{dtc}[DT]{digital twin}
\acrodef{dtn}[DTN]{digital twin network}
\acrodef{fa}[FA]{false alarm}
\acrodef{gnss}[GNSS]{global navigation satellite system}
\acrodef{gps}[GPS]{global positioning system}
\acrodef{imu}[IMU]{inertial measurement unit}
\acrodef{ip}[IP]{incidence point}
\acrodef{las}[L\&S]{localization and sensing}
\acrodef{los}[LOS]{line-of-sight}
\acrodef{mae}[MAE]{mean absolute value}
\acrodef{map}[MAP]{maximum a posteriori}
\acrodef{md}[MD]{miss detection}
\acrodef{mle}[MLE]{maximum likelihood estimator}
\acrodef{mpc}[MPC]{multipath component}
\acrodef{nlos}[NLOS]{non-line-of-sight}
\acrodef{ofdm}[OFDM]{orthogonal frequency division multiplexing}
\acrodef{ransac}[RANSAC]{random sample consensus}
\acrodef{rmse}[RMSE]{root mean square error}
\acrodef{ssb}[SSB]{synchronization signal/physical broadcast channel block}
\acrodef{slam}[SLAM]{simultaneous localization and mapping}
\acrodef{simo}[SIMO]{single-input multiple-output}
\acrodef{tdoa}[TDOA]{time-difference-of-arrival}
\acrodef{toa}[TOA]{time-of-arrival}
\acrodef{ue}[UE]{user equipment}
\acrodef{ura}[URA]{uniform rectangular array}
\acrodef{upa}[UPA]{uniform planar array}
\acrodef{va}[VA]{virtual anchor}

\long\def\comment#1{}

\newfont{\bbb}{msbm10 scaled 700}

\newfont{\bb}{msbm10 scaled 1100}

\renewcommand{\arg}{{\hbox{arg}}}

\begin{document}

\title{Digital Twin-Assisted High-Precision Massive MIMO Localization in Urban Canyons}

\author{Ziqin Zhou$^{\dagger}$, Hui Chen$^{\dagger}$, Gerhard $^{\star}$, Gerhard Steinb\"ock$^{\dagger}$, Henk Wymeersch$^{\star}$ 
\\ 
% \newline \vspace{-3mm} \\
% Please feel free to contact the first author if you have any concerns about the author list and order.

$^{\star}$Chalmers University of Technology, Sweden ~ $^{\dagger}$ 
\vspace{-3mm}
}

\author{Ziqin Zhou\IEEEauthorrefmark{1}, 
Hui Chen\IEEEauthorrefmark{1}, 
Gerhard Steinb\"ock\IEEEauthorrefmark{2}, 
Henk Wymeersch\IEEEauthorrefmark{1}\\
\IEEEauthorblockA{
\IEEEauthorrefmark{1}Department of Electrical Engineering, Chalmers University of Technology, Gothenburg, Sweden 
\\
\IEEEauthorrefmark{2}Ericsson Research, Ericsson, Gothenburg, Sweden %
}
\thanks{This research was supported by the National Growth Fund through the Dutch 6G flagship project {``Future Network Services.''}}
}

\bstctlcite{IEEEexample:BSTcontrol}
\maketitle

\begin{abstract}
High-precision wireless localization in urban canyons is challenged by noisy measurements and severe \ac{nlos} propagation. This paper proposes a robust three-stage algorithm synergizing a \ac{dtc} model with the \ac{ransac} algorithm to overcome these limitations. The method leverages the \ac{dtc} for geometric path association and employs \ac{ransac} to identify reliable \ac{los} and single-bounce \ac{nlos} paths while rejecting multi-bounce outliers. A final optimization on the resulting inlier set estimates the user's position and clock bias. Simulations validate that by effectively turning \ac{nlos} paths into valuable geometric information via the \ac{dtc}, the approach enables accurate localization, reduces reliance on direct \ac{los}, and significantly lowers system deployment costs, making it suitable for practical deployment.
\end{abstract}

\vspace{-3mm}
\acresetall 

\section{Introduction}
A \ac{dt} is a virtual representation that serves as the real-time digital counterpart of a physical object, process, or system \cite{grieves2014digital}. The concept involves three core components: the physical entity, its virtual model, and a continuous data link that ensures the virtual model accurately reflects the state of the physical entity \cite{tao2018digital}. While originating in manufacturing, this paradigm has been widely adopted in other fields, including wireless communications, leading to the vision of a \ac{dt} network \cite{khan2022digital}. A \ac{dt} network creates a virtual replica of the entire communication infrastructure and, crucially, the surrounding radio environment \cite{NDT2022}. A key function of a \ac{dt} network is to act as a physics-based simulation engine that uses its detailed geometric model to simulate complex radio propagation, generating dynamic and accurate radio environment maps on demand \cite{Ruah2024}. This capability is envisioned as a cornerstone for future 6G networks, enabling proactive and predictive services by anticipating future user states and network conditions \cite{zhang20216g}.

Map-aided localization has long been recognized as a powerful approach for improving the accuracy and robustness of wireless positioning systems. Early works leveraged 2D floor plans, often derived from architectural drawings, to impose crucial geometric constraints on the estimation process. For instance, these maps were instrumental in refining the solution space of probabilistic algorithms \cite{Fox1999}. Furthermore, these maps enabled basic \ac{nlos} identification by checking if the direct path between a transmitter and receiver was obstructed by a structure \cite{Djaja2017}. However, the utility of such 2D representations is inherently limited in complex three-dimensional environments like urban canyons, as they fail to capture vertical signal propagation and cannot model the complex multi-path phenomena originating from reflections off building facades.
The advent of high-fidelity 3D models and, more recently, \ac{dt} has opened new frontiers for map-aided localization, offering a much richer description of the environment. Current research on \ac{dt}-aided localization can be broadly categorized into three groups.  First, some approaches treat the \ac{dt} as the ultimate form of a map for \ac{slam}, where an agent builds and refines the map using wireless signals while simultaneously determining its location within it \cite{ge20205g, kaltiokallio2024robust}. A second category of works focuses on leveraging the \ac{dt} for enhanced channel characterization. By generating ``massive fingerprinting'' databases with virtually unlimited density through simulation \cite{morais2403localization}, these methods can better distinguish between \ac{los} and \ac{nlos} conditions, thereby improving accuracy. A third area emphasizes dynamic adaptation, where the DT is continuously synchronized with the physical world to account for changes such as moving objects \cite{Wang2025}. 
Despite these advancements, a critical open challenge remains: effectively leveraging the \ac{dt}'s detailed geometric information to \textit{distinguish and mitigate the effects of severe multi-bounce \ac{nlos} paths}, which are common in complex environments like indoors and urban canyons \cite{kaltiokallio2024robust}. These paths act as geometric outliers that degrade performance, and existing methods often struggle to robustly identify and reject them.

\begin{figure}
    \centering
    \includegraphics[width=0.99\linewidth]{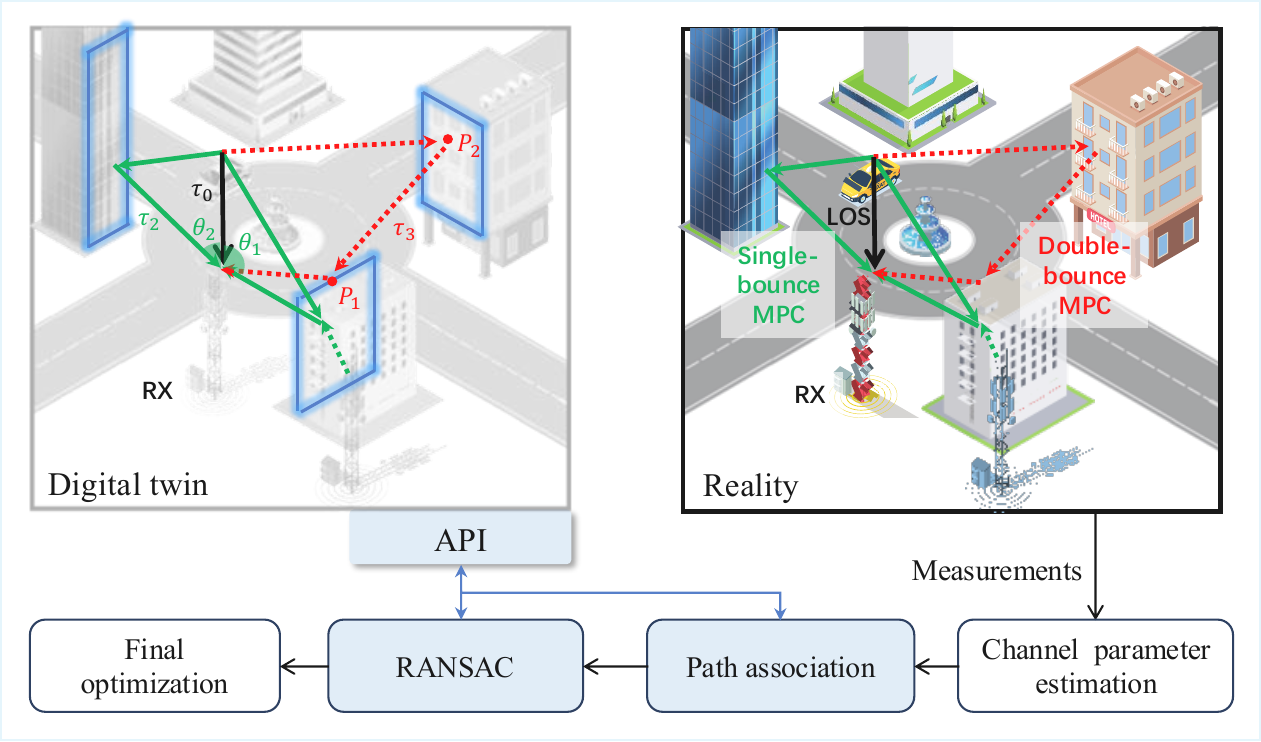}
    \caption{Single-BS localization with aid of a \ac{dt}. The \ac{dt} provides an interface (API) that the localization algorithm can call at various stages (highlighted in blue).}
    \label{fig:scenario} 
\end{figure}

In this paper, we address this challenge by proposing a novel DT-aided localization framework designed to explicitly handle multi-bounce \ac{nlos} paths as geometric outliers. By treating this as an outlier rejection problem, our framework leverages the strong geometric constraints imposed by the \ac{dt} to isolate and filter these inaccurate measurements. The main contributions of this work are as follows:
(i) A novel three-stage robust localization framework that explicitly formulates the multi-bounce \ac{nlos} challenge as a geometric outlier rejection problem. The framework synergizes a \ac{dt} model with the \ac{ransac} algorithm to systematically identify a consensus set of reliable \ac{los} and single-bounce paths from noisy measurements.
(ii) A probabilistic path association technique to robustly generate the initial geometric hypotheses required by the framework. This method explicitly accounts for measurement uncertainty in the \ac{aoa} by using a Monte Carlo-based ray-casting approach, ensuring reliable hypothesis generation even in noisy conditions.
%\end{itemize}

\section{System Model}
As shown in Fig.\ref{fig:scenario}, our system model comprises two distinct yet complementary components: a \ac{dt} model and a generative model. The \ac{dt} serves as a simplified geometric representation of the environment, providing a priori knowledge that the localization algorithm uses to interpret real-world measurements. The generative model, in contrast, is a physics-based simulator designed to mimic the complexity of the real-world measurement process, creating realistic synthetic data. This allows us to use the generative model to generate synthetic data, which serves as the input for our algorithm during validation. %In the real-world application, the algorithm then uses the \ac{dt} directly as a reference map to process live data.

\subsection{Generative Model}

We consider a single-user localization system in a 3D urban environment, modeled using a cartesian coordinate system $[x, y, z]^\top$ with positions in meters. The system comprises a \ac{ue}, located at $\mathbf{u} = [u_x, u_y, u_z]^\top$, and a multi-antenna \ac{bs}, positioned at an elevated location $\mathbf{b} = [b_x, b_y, b_z]^\top$. The \ac{bs} employs a \ac{upa} with $N = N_x \times N_y$ antennas, with an inter-antenna spacing of $d = \lambda/2$, where $\lambda$ is the signal wavelength. 
With limited loss of generality, we consider the \ac{bs} orientation to be aligned with the axes of the coordinate system, avoiding the need to carry the \ac{bs} orientation in the derivations. This alignment allows the measured \ac{aoa} to be directly interpreted as the global \ac{aoa}, which is used throughout this paper. The generative model simulates the wireless signal propagation and subsequent measurements within the environment. The system operates in a \ac{simo} configuration under an \ac{ofdm} scheme. 

\subsubsection{Ray Generation}

For any given \ac{ue} location, the generative model generates $L$ distinct propagation paths between the \ac{ue} and the \ac{bs}, including a potential \ac{los} path and multiple \ac{nlos} reflective paths. Each path $\ell \in \{1, \ldots, L\}$ is characterized by its propagation delay $\tau_\ell$ and \ac{aoa} at the \ac{bs}, which consists of the azimuth angle $\theta_\ell$ and the zenith angle $\phi_\ell$, denoted by the vector $\boldsymbol{\theta}_\ell = [\theta_\ell, \phi_\ell]^\top$. Signal propagation paths are constructed using the principle of specular reflection.
\begin{itemize}
    \item  \textit{\Ac{los} path:} The direct \ac{los} path ($\ell = 1$) is considered valid only if the straight line connecting the \ac{bs} at $\mathbf{b}$ and the UE at $\mathbf{u}$ is not obstructed by any planar surface $S_i$. If the path is unobstructed, its ground-truth delay and \ac{aoa} are calculated as
\begin{align}
    \tau_1 & = \frac{\|\mathbf{u} - \mathbf{b}\|}{c},\\
%\end{align}
%\begin{align}
    \boldsymbol{\theta}_1 &= \Big[ \arctan\Big(\frac{u_y - b_y}{u_x - b_x}\Big), \arccos\Big(\frac{u_z - b_z}{\|\mathbf{u} - \mathbf{b}\|}\Big) \Big]^\top,
\end{align}
where $c$ is the speed of light.

\item \textit{Reflective \ac{nlos} paths:} Specular reflection paths, including both single- and multi-bounce scenarios, are generated using the image method \cite{Yun2015}. As illustrated in Fig.~\ref{fig:scenario}, this technique geometrically determines the point of reflection on a surface, denoted as $\mathbf{r}_\ell$, by creating virtual images of the transmitter (TX) across reflecting surfaces. For instance, the virtual image $TX'$ shown in the figure is used to find the reflection point for the single-bounce path by intersecting the line segment connecting the receiver (RX) and $TX'$ with the surface $S_1$. Multi-bounce paths are found by recursively applying this image creation and intersection process. Each potential path generated this way is then validated for physical feasibility, ensuring that all reflection points $\mathbf{r}_\ell$ lie within their respective surface boundaries and that all path segments remain unobstructed. For every valid path, the total path length yields the ground-truth propagation delay, while the direction of the final segment arriving at the receiver determines the global \ac{aoa}.

\end{itemize}

In addition to these reflections, for simplicity, our generative model is limited to \ac{los} and specular reflection paths. Other propagation effects such as diffuse scattering, diffraction, though supported by modern ray tracers, are not considered in this study.

\subsubsection{Signal Model}
Based on the geometric propagation model described previously, we formulate the uplink signal model of an \ac{ofdm} system. 

For an \ac{ofdm} system with $K$ subcarriers, the \ac{cfr} is given by (for $k\in \{0,1,\ldots,K-1\}$)
%\begin{align}
    $\mathbf{H}[k] = \sum_{\ell=1}^{L} \alpha_\ell \mathbf{a}(\boldsymbol{\theta}_\ell) e^{-j 2\pi f_k \tau_\ell}$,
%\end{align}
where $\alpha_\ell$ is the complex channel gain, $\tau_\ell$ is the propagation delay, $\boldsymbol{\theta}_\ell$ is the \ac{aoa}, $\mathbf{a}(\boldsymbol{\theta}_\ell) \in \mathbb{C}^{N \times 1}$ is the array steering vector, and $f_k$ is the frequency of the $k$-th subcarrier.
Consequently, the received signal vector $\mathbf{y}[k] \in \mathbb{C}^{N \times 1}$ at the \ac{bs} for the $k$-th subcarrier can be written as
\begin{align}
    \mathbf{y}[k] = \mathbf{H}[k] s[k] + \mathbf{n}[k], \label{eq:signalModelOFDM}
\end{align}
where $s[k] \in \mathbb{C}$ is the complex-valued data symbol transmitted by the \ac{ue} on the $k$-th subcarrier. The term $\mathbf{n}[k] \in \mathbb{C}^{N \times 1}$ represents the additive white Gaussian noise (AWGN) vector, which is modeled as a circularly-symmetric complex normal random variable, i.e., $\mathbf{n}[k] \sim \mathcal{CN}(\mathbf{0}, \sigma^2 \mathbf{I}_N)$, with $\sigma^2$ being the noise variance and $\mathbf{I}_N$ being the $N \times N$ identity matrix.

\subsection{Measurement Model}

In a practical system, the \ac{bs} would estimate the delay and \ac{aoa} for each path by processing uplink reference signals from the \ac{ue} using standard channel paramter estimation methods such as sparsity-based methods (e.g., OMP\cite{OMP}) subspace-based methods (e.g., MUSIC\cite{MUSIC}). This work focuses on the positioning part and assumes the channel parameters are obtained from \eqref{eq:signalModelOFDM} using an efficient estimator, resulting in the following measurement model. % to be described next. 
% Our generative model simulates this process by adding zero-mean Gaussian noise to the ground-truth channel parameters derived from ray generation.
The measured propagation delay for path $\ell$  is modeled as
%\begin{align}
    $\hat{\tau}_\ell = {L_\ell}/{c} + B + n_{\tau_\ell}$, 
%\end{align}
where the path length $L_\ell$ is $\|\mathbf{u} - \mathbf{b}\|$ for the \ac{los} path. For \ac{nlos} paths, it is the sum of the lengths of all constituent segments (e.g., $\|\mathbf{u} - \mathbf{r}_\ell\| + \|\mathbf{r}_\ell - \mathbf{b}\|$ for a single-bounce path). The model includes an unknown clock bias $B$ common to all paths and path-specific measurement noise $n_{\tau_\ell} \sim \mathcal{N}(0, \sigma_{\tau_\ell}^2)$.
The measured \ac{aoa} is modeled by 
%\begin{align}
   $ \hat{\boldsymbol{\theta}}_\ell = \boldsymbol{\theta}_\ell + \mathbf{n}_{\boldsymbol{\theta}_\ell}$, 
%\end{align}
where $\mathbf{n}_{\boldsymbol{\theta}_\ell} = [n_{\theta_\ell}, n_{\phi_\ell}]^\top$ is the measurement noise vector. Its components are modeled as independent, zero-mean Gaussian\footnote{More generally, a Von Mises model could be used.} random variables with potentially different variances, i.e., $n_{\theta_\ell} \sim \mathcal{N}(0, \sigma_{\theta}^2)$ and $n_{\phi_\ell} \sim \mathcal{N}(0, \sigma_{\phi}^2)$.

\subsection{Digital Twin}\label{dtsection}
The \ac{dt} is a digital representation of the urban environment, implemented as a functional object that encapsulates the geometry of a set of $M$ planar surfaces, $\{S_i\}_{i=1}^M$. Each surface $S_i$ is defined by key geometric properties, including a unit normal vector $\mathbf{n}_i = [n_{ix}, n_{iy}, n_{iz}]^\top$, a point on the surface $\mathbf{p}_i = [p_{ix}, p_{iy}, p_{iz}]^\top$, and its finite boundaries. The \ac{dt} provides a well-defined interface with three primary methods for geometric queries.

The first method, $i = \texttt{FindSurface}(\mathbf{d},\mathbf{b})$, serves as a fundamental ray-casting tool. Given a ray originating from the \ac{bs} at position $\mathbf{b}$ with a direction vector $\mathbf{d}$, its purpose is to identify the first environmental surface the ray intersects. The implementation finds the intersection point $\mathbf{r} = \mathbf{b} + t\mathbf{d}$ with the minimum positive distance $t$ for every surface in the \ac{dt}. If this point lies within the surface's finite boundaries, the method returns the index of that surface.

The function $\mathbf{r} = \texttt{IncidencePoint}(\mathbf{u},\mathbf{b},i)$ calculates the exact 3D coordinates of a specular reflection on a given surface $S_i$. It takes as input the positions of the \ac{ue} at $\mathbf{u}$ and the \ac{bs} at $\mathbf{b}$. The method's principle is the image method, where it first computes the \ac{ue}'s virtual image, $\mathbf{u}'$, with respect to the plane of surface $S_i$
\begin{align}
    \mathbf{u}' = \mathbf{u} - 2 ((\mathbf{u} - \mathbf{p}_i)^\top \mathbf{n}_i) \mathbf{n}_i.
\end{align}
The reflection point $\mathbf{r}$ is then found at the intersection of the line segment connecting the \ac{bs} and this virtual image with the plane of $S_i$. This is computed by solving for the intersection parameter of the line connecting $\mathbf{b}$ and $\mathbf{u}'$
\begin{align}
    \mathbf{r} = \mathbf{b} + \frac{(\mathbf{p}_i - \mathbf{b})^\top \mathbf{n}_i}{(\mathbf{u}' - \mathbf{b})^\top \mathbf{n}_i} (\mathbf{u}' - \mathbf{b}).
\end{align}
The function outputs the coordinates of this point, provided that $\mathbf{r}$ falls within the finite physical boundaries of the surface.

\subsection{Problem Formulation}
The objective of this work is to jointly estimate the UE position $\mathbf{u}$ and the clock bias $B$, from the observations $\{\mathbf{y}[k]\}_{k=0}^{K-1}$, leveraging the \ac{dt}. %The observations are converted to $L$ uplink measurements $\{\hat{\tau}_\ell, \hat{\boldsymbol{\theta}}_\ell\}_{\ell=1}^L$ with associated variances and a \ac{dt} of the environment, find the state vector $(\mathbf{u}, B)$ that best explains these measurements. 
Note that without the \ac{dt}, the \ac{ue} cannot be localized. Hence, the \ac{dt} must be used to identify single-bounce \ac{nlos} paths and identify them to known surfaces, to render the localization problem identifiable.

\section{\ac{dt}-aided Localization Method}

To estimate the \ac{ue} position $\mathbf{u}$ and clock bias $B$ in the presence of noisy measurements and multi-bounce \ac{nlos} paths, we propose a robust, three-stage methodology. 
%\begin{enumerate}
 %   \item 
First, we perform a candidate generation step to establish preliminary measurement-to-surface associations. In this stage, we generate a set of path hypotheses $\{h_\ell\}$, where each hypothesis $h_\ell$ is a tuple $(\ell, i)$ that pairs a measurement index $\ell$ with a hypothesized geometric origin index $i$ (i.e.,  surface $S_i$) of the \ac{dt}. 
%\item 
Second, we apply the \ac{ransac} algorithm to this expanded set of hypotheses to robustly identify a consensus set corresponding to the true LOS and single-bounce paths (inliers), while filtering out those from unmodeled multi-bounce paths (outliers). 
%\item 
Third, we perform a refined optimization using the maximum likelihood criterion using only the final inlier set to obtain accurate estimates of $\mathbf{u}$ and $B$.
%\end{enumerate}
These three stages are elaborated in the following sections.

\subsection{Path Association}
For any given path measurement \(\ell \in \{1, \dots, L\}\), we  generate a hypothesis for its origin (i.e., the associated surface in the \ac{dt}). Leveraging the strong prior that the \ac{los} path exhibits the minimum travel time, we identify the path with the minimum measured delay, $\ell^* = \arg\min_\ell \hat{\tau}_\ell$, and assign it the \ac{los} hypothesis, $h_{\ell^*} = 0$. For all other paths, the goal is to identify the most likely reflecting surface from the set \(\{S_1, \dots, S_M\}\) to assign an \ac{nlos} hypothesis, $h_\ell $. 
The association process leverages the interface of the \ac{dt}. To account for the measurement noise, for each \ac{nlos} path \(\ell\), we generate a set of \(K\) perturbed \ac{aoa}-based directions (denoted by \(\{\mathbf{d}_\ell^{(k)}\}_{k=1}^K\)) by sampling from a Gaussian distribution centered at the measured \ac{aoa} as 
%\begin{align}
    $\theta_\ell^{(k)} \sim \mathcal{N}(\hat{\theta}_\ell, \sigma_{\theta}^2)$ and $\phi_\ell^{(k)} \sim \mathcal{N}(\hat{\phi}_\ell, \sigma_{\phi}^2)$,
%\end{align}
leading to 
%\begin{align}
    $\mathbf{d}^{(k)}_\ell = [\sin{\phi}^{(k)}_\ell \cos{\theta}^{(k)}_\ell, \sin{\phi}^{(k)}_\ell \sin{\theta}^{(k)}_\ell, \cos{\phi}^{(k)}_\ell]^\top$.
%\end{align}

We then query the \ac{dt} with this direction as $i^{(k)}_\ell = \texttt{FindSurface}(\mathbf{d}^{(k)}_\ell,\mathbf{b})$, which returns the index of the intersected surface.
We then assign a score to each surface
\begin{align}
\text{Score}_\ell(i) = \frac{1}{K} \sum_{k=1}^K \mathbb{I}({i_\ell^{(k)}} = i),
\end{align}
where \(\mathbb{I}(\cdot)\) is the indicator function. The surface with the highest score is selected as the associated surface, 
\begin{align}
 h_\ell &  = \arg\max_i \text{Score}_\ell(i)\\
 & \text{s.t.} \,\, \text{Score}_{\ell}(i)>\gamma, \notag
\end{align}
where $\gamma$ is a threshold, to filter out paths that lack a clear geometric origin.

\subsection{Path Classification Using RANSAC}
The second stage of our methodology employs the \ac{ransac} algorithm to robustly classify the initial path hypotheses. It iteratively generates candidate states \((\mathbf{u}, B)\) from minimal subsets of measurements and selects the state that is consistent with the largest consensus set of inliers (\ac{los} and single-bounce paths). Multi-bounce paths, which are inconsistent with the model, are discarded as outliers.

The core of this classification process is the residual cost function, $f_\ell(\mathbf{u}, B, h_\ell)$, which quantifies the agreement between a measurement and a given hypothesis. Assuming independent Gaussian noise, this cost is derived from the negative log-likelihood
\begin{align}\label{fl}
&f_\ell(\mathbf{u}, B, h_\ell) \triangleq \frac{1}{\sigma_{\tau_\ell}^2} \Big( \tilde{\tau}_\ell - \frac{d_\ell(\mathbf{u}, h_\ell)}{c} - B \Big)^2 
+ \\&\frac{1}{\sigma_{\theta_\ell}^2} \Big( \tilde{\theta}_\ell - \theta_\ell(\mathbf{u}, h_\ell) \Big)^2 \notag + \frac{1}{\sigma_{\phi_\ell}^2} \Big( \tilde{\phi}_\ell - \phi_\ell(\mathbf{u}, h_\ell) \Big)^2.
\end{align}

The prediction functions $d_\ell(\mathbf{u}, h_\ell)$  and $(\theta^{\text{pred}}_\ell, \phi^{\text{pred}}_\ell)$ within this cost model compute ideal path length and \ac{aoa} based on the hypothesis $h_\ell$. The calculation relies on the \ac{dt}'s \texttt{IncidencePoint} function to determine the reflection point for \ac{nlos} paths, while \ac{los} parameters are computed directly.

The \ac{ransac}-based  classification\cite{kaltiokallio2024robust} proceeds as follows:
\begin{itemize}
    \item \textit{Random sampling}: A minimal subset \(\mathcal{S}\) from \(L\) paths is randomly selected. We set \(|\mathcal{S}|=2\) to form a solvable system for the four unknown state variables \((\mathbf{u}, B)\). 

    \item \textit{Model estimation}: For the selected hypothesis, a candidate state \((\mathbf{u}, B)\) is computed by minimizing the cost function over the minimal set \(\mathcal{S}\): %using the same optimization method as described in Section~\ref{optimize}
    \begin{align}\label{P1}
    \min_{\mathbf{u}, B} \sum_{\ell \in \mathcal{S}} f_\ell(\mathbf{u}, B, h_\ell),
    \end{align}
    subject to geometric constraints, where the cost function $f_\ell$ is defined as in (\ref{fl}).

    \item \textit{Inlier Evaluation}: The estimated state \((\mathbf{u}, B)\) is used to find the consensus set by re-evaluating all paths, \(\ell \in \{1, \dots, L\}\). Each path is individually classified as an inlier if its cost, \(f_\ell(\mathbf{u}, B, h_\ell)\), does not exceed a predefined threshold \(T\). This step ensures that unmodeled paths, such as multi-bounce reflections, are correctly rejected as outliers due to their typically large cost.

    \item \textit{Iteration and selection}: The process is repeated for \(N\) iterations, where \(N\) is calculated to ensure a high probability (\(p\)) of selecting an outlier-free subset by 
    $N = {\log(1 - p)}/{\log(1 - (1 - \epsilon)^s)}$
 \cite{ransac1981}, 
 %   \begin{align}
  %      N = \frac{\log(1 - p)}{\log(1 - (1 - \epsilon)^s)},
  %  \end{align}
    with \(\epsilon\) as the expected outlier ratio. The model \((\mathbf{u}, B)\) that yields the largest set of inliers, denoted \(\mathcal{L}_{\text{inlier}}\), is chosen. %Any ties in the number of inliers are resolved by selecting the model with the minimum sum of costs over its respective inlier set\cite{kaltiokallio2024robust}.
\end{itemize}

\subsection{Final Optimization}\label{optimize}
Given the final inlier set \(\mathcal{L}_{\text{inlier}}\) and their confirmed path types (\ac{los} or \ac{nlos} with a specific surface), we refine the estimates of \(\mathbf{u}\) and \(B\) by solving the full maximum likelihood problem over all inliers:
\begin{align}
\min_{\mathbf{u}, B} \sum_{\ell \in \mathcal{L}_{\text{inlier}}} f_\ell(\mathbf{u}, B, h_\ell),
\end{align}
subject to the position bounds \(x_u, y_u \in [x_{\min}, x_{\max}], z_u \in [0, z_{\max}]\).  The reflection point \(\mathbf{r}_\ell\) for each \ac{nlos} path is implicitly determined by the candidate position \(\mathbf{u}\) and its known reflecting surface. This problem is a standard non-linear least squares problem and is typically solved using an iterative solver, such as Levenberg-Marquardt.

To ensure robust and fast convergence, a well-chosen initial point is critical. Instead of constructing a separate geometric initial point, we leverage the output from the RANSAC stage itself. The RANSAC algorithm inherently identifies the candidate state $(\mathbf{u}, B)$ that is consistent with the largest consensus set. This state, having been generated from a minimal (and likely outlier-free) sample and validated by the maximum number of inliers, serves as an excellent and robust initial estimate. Therefore, we initialize this final optimization using the state vector obtained from the winning RANSAC iteration.

\section{Simulation Results}

\subsection{Simulation Setup and Environment Configuration}
The performance of the proposed algorithm is evaluated through simulations. Unless otherwise specified, the key parameters are set as follows: the receiver employs an $8 \times 8$ antenna array with an element spacing of half a wavelength, the number of OFDM subcarriers is 512, the subcarrier spacing is 30~kHz, the center frequency is 3.5~GHz, and the receiver noise figure is 7~dB. These parameters are crucial as they determine the theoretical estimation accuracy of the channel parameters. We leverage the Cram\'{e}r-Rao lower bound (CRLB) to model this relationship; specifically, the measurement noise variances ($\sigma_{\tau_\ell}^2, \sigma_{\theta_\ell}^2, \sigma_{\phi_\ell}^2$) used in our simulations are set according to the CRLB, although the detailed derivation is omitted for brevity.

The simulation scenario is set within a specific urban canyon-like environment, geometrically represented by a \ac{dt} model that includes three primary reflecting surfaces. This DT model serves as the a priori environmental knowledge for our localization algorithm. 

Separately, to generate the ground truth for our simulations, we consider the signal propagation within this same environment. For the specific placement of the \ac{bs} at $(0, 0, 15)$~m and the \ac{ue} at $(-15, -15, 0)$~m, the propagation results in seven dominant multipath components. These ground-truth paths, used as the basis for generating noisy measurements, comprise one \ac{los} path, three single-bounce \ac{nlos} paths whose reflections correspond geometrically to the surfaces defined in the DT, and three multi-bounce \ac{nlos} paths.

\subsection{Performance Metric}
We evaluate the performance of the proposed algorithm at different stages using several key metrics. First, for the initial path hypothesis generation stage, we use the correct association rate, defined as the percentage of true single-bounce \ac{nlos} paths that are correctly associated with their ground-truth reflecting surface. Subsequently, to assess the \ac{ransac} classification stage, we employ two standard metrics: the \ac{fa} Rate, which measures the proportion of multi-bounce \ac{nlos} paths (true outliers) incorrectly classified as inliers, and the \ac{md} Rate, which measures the proportion of true \ac{los} or single-bounce \ac{nlos} paths (true inliers) incorrectly classified as outliers. All these rates are averaged over $N_{MC}$ Monte Carlo runs, where each run involves generating a new random realization of the measurement noise ($n_{\tau_\ell}$ and $\mathbf{n}_{\boldsymbol{\theta}_\ell}$) added to the true path parameters for all seven paths. Finally, the overall localization accuracy is evaluated using the \ac{rmse}, which quantifies the Euclidean distance between the estimated \ac{ue} position $\hat{\mathbf{u}}$ and the ground-truth position $\mathbf{u}$, also averaged over the runs:
\begin{align}
    \text{RMSE} = \sqrt{\frac{1}{N_{ \text{MC}}} \sum_{n=1}^{N_{\text{MC}}} \|\hat{\mathbf{u}}^{(n)} - \mathbf{u}\|^2},
\end{align}
where $\hat{\mathbf{u}}^{(n)}$ is the position estimate from the $n$-th run. Unless otherwise specified, all metrics presented below are averaged over 500 Monte Carlo runs.

\subsection{Path Association Performance}
\begin{figure}
    \includegraphics[width= 0.95\columnwidth]{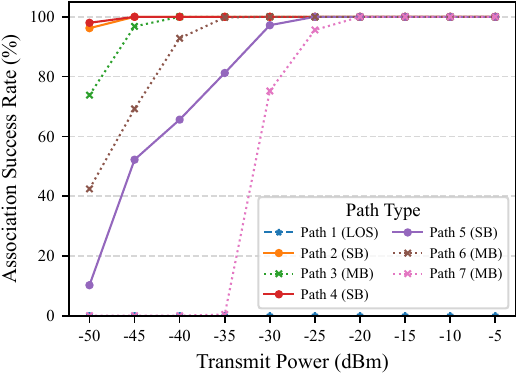}
    \caption{Path association success rate versus transmit power (SB: Single-Bounce, MB: Multi-Bounce)}
    \label{fig:PA}
\end{figure}

Fig.~\ref{fig:PA} evaluates the performance of the path association stage by plotting the correct association rate for NLOS paths versus transmit power. 

As expected, the correct association rate improves with higher transmit power. This is because increased power leads to lower measurement noise variance ($\sigma_{\theta}^2, \sigma_{\phi}^2$), which in turn causes the $K=1000$ perturbed rays in our probabilistic casting method to be more tightly clustered around the true direction of arrival. This increases the likelihood that the score for the correct surface will be the maximum and will exceed the confidence threshold of $\gamma=0.7$. It is important to note that this evaluation focuses exclusively on the performance for NLOS paths. The LOS path is not part of this test, as it is handled in a separate, deterministic step by identifying the path with the minimum delay.\footnote{We acknowledge that the minimum-delay prior can fail under very low SNR conditions, where a true \ac{nlos} path might be misidentified as the \ac{los} candidate. Our methodology is inherently robust to this potential misclassification. The subsequent \ac{ransac} stage employs a uniform random sampling strategy, which ensures a non-zero probability of selecting a minimal subset composed exclusively of true single-bounce \ac{nlos} paths. A model generated from such a 'pure' sample will be highly accurate and will, in turn, correctly identify the mislabeled minimum-delay path as a high-cost outlier. Therefore, the overall algorithm's success does not depend on the initial \ac{los} hypothesis being correct.} The figure shows that when the transmit power exceeds -20 dBm, the correct association rate for all true NLOS paths approaches 100\%, demonstrating the robustness of our proposed method. 

\subsection{RANSAC Performance}
First, we evaluate the filtering performance of the RANSAC algorithm using \ac{fa} and \ac{md} rates. 

\begin{figure}
    \includegraphics[width=0.95\columnwidth]{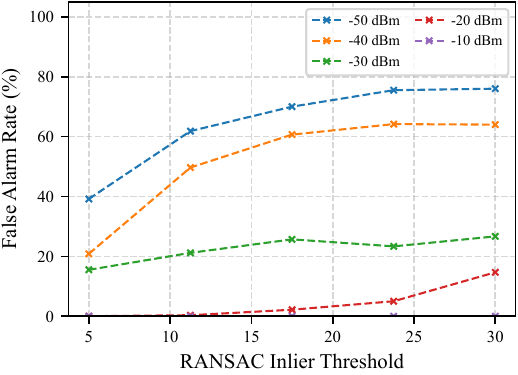}
    \caption{\ac{fa} of RANSAC with different threshold versus transmit power}
    \label{fig:FA}
\end{figure}

Fig.~\ref{fig:FA} illustrates the \ac{fa} rate of the \ac{ransac} classification stage. While \ac{fa} are observed, particularly at lower transmit power levels, the \ac{fa} rate decreases rapidly as power increases, demonstrating the algorithm's effectiveness in rejecting outliers under reasonable signal conditions.

This behavior stems from the interplay between geometric inconsistency and measurement uncertainty. At very low transmit power, the large measurement uncertainty (high noise variance) can obscure the inherent geometric mismatch between a multi-bounce path and the single-bounce model in our cost function. This allows an outlier to fortuitously appear consistent with a candidate model, resulting in a false alarm. However, as transmit power increases, the measurement uncertainty diminishes significantly. The underlying geometric mismatch then dominates, leading to a high residual cost for multi-bounce paths and their correct rejection as outliers. Consequently, the algorithm achieves a very low \ac{fa} rate once sufficient transmit power is available.

\begin{figure}
    \includegraphics[width=0.95\columnwidth]{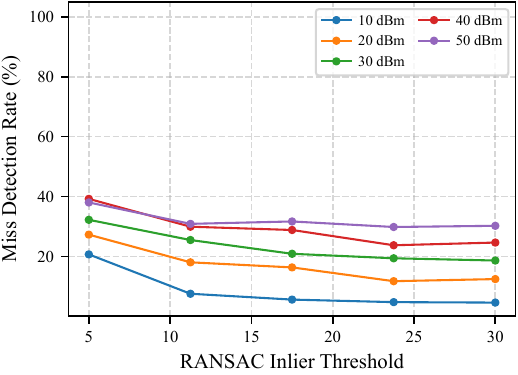}
    \caption{\ac{md} of RANSAC with different threshold versus transmit power}
    \label{fig:MD}
\end{figure}

Fig.~\ref{fig:MD} illustrates the \ac{md} rate. Counter-intuitively, the results show that the MD rate begins to increase as the transmit power becomes very high. This phenomenon can be attributed to the cost function's hypersensitivity to model inaccuracies at low noise levels.

The cost, $f_\ell$, is a squared error normalized by the measurement variance ($\sigma_{\tau_\ell}^2, \sigma_{\theta_\ell}^2, \sigma_{\phi_\ell}^2$), which shrinks towards zero as transmit power increases. In any given RANSAC iteration, the candidate state \((\mathbf{u}, B)\) is estimated from a minimal, random sample and is therefore only an approximation of the true state. Consequently, when a true inlier is evaluated against this slightly imperfect model, the small but non-zero geometric mismatch between the measurement and the prediction results in an extremely large normalized cost $f_\ell$. This ``cost explosion'' causes the value to exceed the fixed inlier threshold $T$, leading to the erroneous rejection of a true inlier. As expected, the figure also confirms that a higher threshold $T$ mitigates this effect by providing a larger acceptance region.

\subsection{Localization Performance}
Finally, we evaluate the localization performance under various \ac{ransac} thresholds and transmission powers, using the \ac{rmse} as the metric.

\begin{figure}
    \includegraphics[width=0.95\columnwidth]{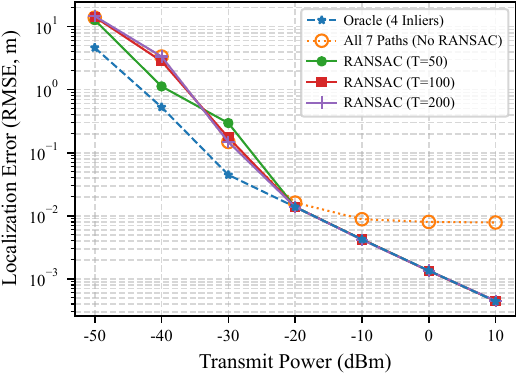}
    \caption{\ac{rmse} with different threshold \ac{ransac} versus transmit power}
    \label{fig:locerr}
\end{figure}

As shown in Fig.~\ref{fig:locerr}, the localization error generally decreases as transmit power increases. Compared to the oracle performance using only true inliers ('Perfect Inlier'), simply using all available paths without outlier rejection ('All Paths') performs the worst, with its error saturating at a high value due to the bias introduced by multi-bounce outliers. In contrast, the \ac{ransac} approach significantly improves accuracy, although performance varies with the threshold $T$. Notably, the localization performance is optimal across all tested thresholds within the transmit power range of {$-$20~dBm} to 10~dBm. This optimal window aligns well with the performance observed in the preceding stages: path association achieves near-perfect success, the \ac{fa} rate effectively vanishes, and the \ac{md} rate remains low within this regime.

\section{Conclusion}
This paper presented a \ac{dt}-assisted high-precision localization framework for massive MIMO systems operating in complex urban environments. By leveraging the detailed geometric knowledge provided by the \ac{dt}, the proposed method effectively associates signal paths with physical surfaces and removes multi-bounce \ac{nlos} outliers. The final optimization over the inlier set enables accurate estimation of both user position and clock bias. Simulation results in a 3D scenario demonstrated the effectiveness of the \ac{ransac}-based outlier rejection algorithm, achieving near-oracle accuracy with significantly reduced deployment complexity. Future work will extend this framework toward dynamic and large-scale environments, where the updating of \ac{dt} will also be considered.

\bibliographystyle{IEEEtran}
\bibliography{ref}

\vfill

\end{document}